# Atomic-scale insights into semiconductor heterostructures: from experimental three-dimensional analysis of the interface to a generalized theory of interface roughness scattering

T. Grange[1,§], S. Mukherjee[2,§], G. Capellini[3,4,*], M. Montanari[4], L. Persichetti[4], L. Di Gaspare[4], S. Birner[1], A. Attiaoui[2], O. Moutanabbir[2], M. Virgilio[5], and M. De Seta[4]

[1] nextnano GmbH, Garching b. München D-85748, Germany

[2] Department of Engineering Physics, École Polytechnique de Montréal, Canada

[3] IHP – Leibniz-Institut für innovative Mikroelectronik, Im Technologiepark 25, D-15236 Frankfurt (Oder), Germany

[4] Dipartimento di Scienze, Università degli Studi Roma Tre, I-00146, Roma, Italy

[5] Dipartimento di Fisica "E. Fermi," Università di Pisa, Largo Pontecorvo 3, I-56127 Pisa, Italy

* email to: capellini@ihp-microelectronics.com

## ABSTRACT

In this manuscript, we develop a generalized theory for the scattering process produced by interface roughness on charge carriers and which is suitable for any semiconductor heterostructure. By exploiting our experimental insights into the three-dimensional atomic landscape of Ge/GeSi heterointerfaces obtained by atom probe tomography, we have been able to define the full set of interface parameters relevant to the scattering potential, including both the in-plane and axial correlation inside real diffuse interfaces. Our experimental findings indicate a partial coherence of the interface roughness along the growth direction within the interfaces. We show that it is necessary to include this feature, previously neglected by theoretical models, when heterointerfaces characterized by finite interface widths are taken into consideration. To show the relevance of our generalized scattering model in the physics of semiconductor devices, we implemented it in a non-equilibrium Green's function simulation platform to assess the performance of a Ge/SiGe-based THz quantum cascade laser.

[§] *these authors contributed equally to this work*

## I. INTRODUCTION

Nowadays, the boundary between fundamental research and technology in the field of semiconductor science is becoming more and more blurred, as witnessed by the increasingly shorter time required to bring innovative materials and processes to mass production [1]. As an example, "*exotic*" quantum effects can now be exploited in nanoscale devices in order to improve their performance or to add functionalities. In this scenario, the interface in semiconductor heterostructures plays an ever-greater role and its characterization down to the atomic scale is becoming more relevant. Indeed, a well-defined and sharp interface is a prerequisite for a number of different physical systems leveraging on 2D effects, such as solid state based qubits [2], thermoelectric devices [3, 4], photonic devices [5, 6], and also for more conventional architectures as field-effect transistors [7-9], to name a few. Furthermore, in the sub-10 nm CMOS technology nodes, interfaces are critical in developing high quality nano-sheet channels [1].

The interdiffusion of atoms and the 2D/3D island formation [10] makes a *real* heterointerface deviating from an *ideal* flat atomic plane in two important aspects: (*i*) the average composition profile is smeared out along the growth (axial) direction, known as interface broadening and usually quantified by the interface width $\mathcal{L}$; (*ii*) the iso-compositional surfaces exhibit in-plane fluctuations of their axial position, which are responsible for the so-called interface roughness (IFR).

The interface broadening and roughness influence the opto-electronic properties in different ways and thus, they are usually modelled separately. The interface broadening results in the broadening of the potential experienced by charge carriers in the *z* direction, *i.e.* perpendicular to the interface plane. This can be accounted for by replacing the *ideal* 1D, box-like (abrupt) potential profile with the *real* broadened *z*-dependent potential profile in the one-dimensional Schrödinger equation. In many situations, electronic wave functions and energies are not severely affected by smeared-out interfaces; thus their effect can be neglected or easily taken into account when designing the active layer region [11, 12]. By contrast, the interface roughness, which breaks the in-plane invariance, has to be treated as a perturbing potential term responsible for scattering among different eigenstates [13-15]. This scattering channel has been extensively modeled only for abrupt interfaces [13, 14, 16], with the roughness described as an in-plane $z(x,y)$ fluctuation of the interface position. This fluctuation distribution is typically characterized by its root mean square (rms) $\Delta$ and in-plane correlation length $\Lambda_\parallel$. Recently, Valavanis *et al*. [17] generalized the IFR modelling to treat diffuse interfaces as well. Nevertheless, this model is limited to the case of a perfectly correlated roughness along the growth direction, implying that the out-of-plane fluctuations of all the iso-composition surfaces, defining the diffused interface region at each (*x,y*) point, are equaled. It is clear that this approximation is not valid when the correlation length of the fluctuations along $z$ (axial correlation length $\Lambda_\perp$) is comparable or smaller than the interface width $\mathcal{L}$.

Pulsed laser-assisted atom probe tomography (APT) has recently emerged as the ideal technique for evaluating the interface width $\mathcal{L}$ [18]. Furthermore, APT makes it possible to reconstruct atom-by-atom the crystal structure in three dimensions and, potentially, to quantify other properties of buried interfaces, such as the rms roughness $\Delta$, the in-plane correlation length [19] $\Lambda_\parallel$, and the axial correlation length $\Lambda_\perp$.

Motivated by the recent interest in Ge/SiGe-based devices [20-22], here we show how it was possible to measure all the four interface parameters ($\mathcal{L}$, $\Delta$, $\Lambda_\parallel$, and now $\Lambda_\perp$ as well), in a sample consisting of a stack of strain-compensated Ge/Ge$_{0.8}$Si$_{0.2}$ asymmetric coupled quantum wells (ACQWs). Our data demonstrate the excellent interface quality in terms of small interface width ($\mathcal{L}$ =1.16 nm) and interface rms roughness ($\Delta$=0.18 nm). Interestingly we found that the interface roughness is vertically correlated along the growth axis on a length $\Lambda_\perp$ =0.26 nm smaller than the interfacial width. This outcome prompted us to develop a general theory for interface roughness scattering, accounting for both the finite correlation length along the growth direction and the interference effects associated with the interaction among different interfaces. We show that the effect of a finite axial correlation length has a large impact on the IFR scattering rate and that knowledge of the IFR at the atomic scale and its proper modelling are essential when assessing the performance of quantum devices. Indeed, we incorporated our findings into a non-equilibrium Green's function (NEGF) simulation platform [21] to model the optoelectronic properties of an n-type Ge/SiGe THz QCL. We chose this device, which can be considered a stack of different ACQWs, as the ideal case study for assessing the relevance of our generalized model, given that heterointerfaces play a very important role in its operating principle. We found that the finite axial correlation length critically affects the QCL gain. Interestingly, for the set of interface parameters ($\mathcal{L}$, $\Delta$, $\Lambda_\parallel$, $\Lambda_\perp$) which we measured in our sample, we can predict a room temperature operation of the device.

**II. METHODS**

The Ge/Ge$_{0.8}$Si$_{0.2}$ ACQW sample was grown by ultra-high-vacuum chemical vapor deposition (UHV-CVD) at 500 °C, using germane and silane, without carrier gases, on Si(001) substrates. The ACQW stack was deposited on top of a SiGe reverse-graded (RS) virtual substrate (VS), ending with a 1.5 µm-thick constant composition fully relaxed Ge$_{0.87}$Si$_{0.13}$. To avoid plastic relaxation accompanied by lattice defects in the ACQW region, the composition of the RG-VS was chosen to achieve strain compensation condition in the ACQW, featuring tensile strained Ge$_{0.8}$Si$_{0.2}$ barriers and compressively strained Ge well layers. Each of the 20 modules has been designed to have a thick ($d_{tw}$ =12 nm) and a narrow ($d_{nw}$ = 5 nm) Ge well, coupled to a $d_{tb}$=2.3 nm Ge$_{0.8}$Si$_{0.2}$ tunnel barrier. Another $d_{sb}$=20 nm-thick Ge$_{0.8}$Si$_{0.2}$ layer is used to separate each module, so that the nominal period thickness of the structure is 39.3 nm. More details about the growth and X-ray diffraction (XRD) structural characterization of ACQW samples are reported in Ref. [23]. The sample preparation for the high-resolution scanning transmission electron microscopy (HR-STEM) analysis and the APT analysis was done using a Helios Nanolab 650 dual channel (Ga$^+$ ion column for milling, SEM column for imagining) focused-ion beam (Dual-FIB) microscope, using the standard lamella lift-out technique. The HR-STEM analysis was conducted using a double cross-section-corrected FEI Titan microscope, operated at 200 kV. The beam converge angle was 19.1 mrad. Aberration corrected magnetic lenses helped in making the probes of the order of 1-2 Å diameter with a beam currents of 200 pA. CEOS CESCOR corrector was used to yield a resolution of 0.8 Å. The images were recorded using a high-angle annular dark field (HAADF) detector and the data were processed using the digital micrograph GMS3 software. As for the APT experiments, a stable and layer-by-layer evaporation of atoms was achieved focusing a picosecond-pulsed ultraviolet laser on the FIB defined nanotips. The laser energy, pulse frequency, and the evaporation rate (ions/pulse) were maintained at 5.0-10.0 pJ, 250-500 kHz, and 0.5-1.5, respectively. The constant shank-angle based 3-D reconstructions were performed using the Cameca's integrated visualization and analysis software (IVAS) package. In an APT set-up, atoms are evaporated as cations, in layer-by-layer fashion, starting from the apex of the tip, while the software re-arranges these atoms in the same sequence as they are evaporated (detected) to create the 3D atomistic reconstruction.

The NEGF simulations are performed using the nextnano.QCL package, which calculates the quantum transport through layered heterostructures. The scattering processes are treated within the self-consistent Born approximation, including phonons, charged impurities, electron-electron, interface roughness, and alloy disorder [21, 24, 25]. We neglected hot-phonon effects assuming a Bose-Einstein statistics at the lattice temperature. Moreover, aiming at reducing the computational time, the electron-electron interaction is calculated exploiting a mean-field approximation, as discussed in Ref.[21], where the interest reader can find a detailed description of the model and of its approximations. The gain is calculated self-consistently within the linear response theory following Ref. [26]. The graded composition profile of the QCL is calculated by convoluting the nominal squared profile with a Gaussian function of full-width at half maximum (FWHM) $\mathcal{L}$ [11]. As detailed in the Supplementary Material (SM) [27], the interface roughness self-energy is derived according to the present generalized theory of interface roughness.

## III. RESULTS AND DISCUSSION

### A. Experimental determination of interface parameters in Ge/SiGe ACQWs

The HAADF-STEM image acquired on the ACQWs stack is shown in Fig. 1(a). Images recorded at progressively higher magnifications have been highlighted by color-coded rectangular boxes and are displayed as insets. We can clearly distinguish the stacking along the growth direction of 20 identical periods separated by a $Ge_{0.8}Si_{0.2}$ layer (darker grey). Lines scans (averaged over 100 pixels) were recorded using the GSM3 software in order to extract the intensity variation in the HAADF-STEM images across the entire thickness of the ACQW stack. We measured the mean thicknesses of the wells and barriers by fitting the intensity profile with Gaussian functions (not shown) to be $d_{sb}$ =19.60 nm ($\pm$1.3 %) for the SiGe spacer barrier, $d_{tw}$ = 12.22 nm ($\pm$1.4 %) and $d_{nw}$ = 5.15 nm ($\pm$1.2 %) for the thick and narrow Ge well respectively, and $d_{tb}$ = 2.48 nm ($\pm$1.2 %) for the SiGe tunneling barrier, resulting in an average total period thickness of 39.5 nm ($\pm$5.1%) (the uncertainties represent the standard deviation of the thicknesses measured over the 20 periods of the ACQW). HAADF-STEM analysis has confirmed, within the experimental uncertainty, the thickness reproducibility of all the individual layers along the stack and the good match to the nominal values.

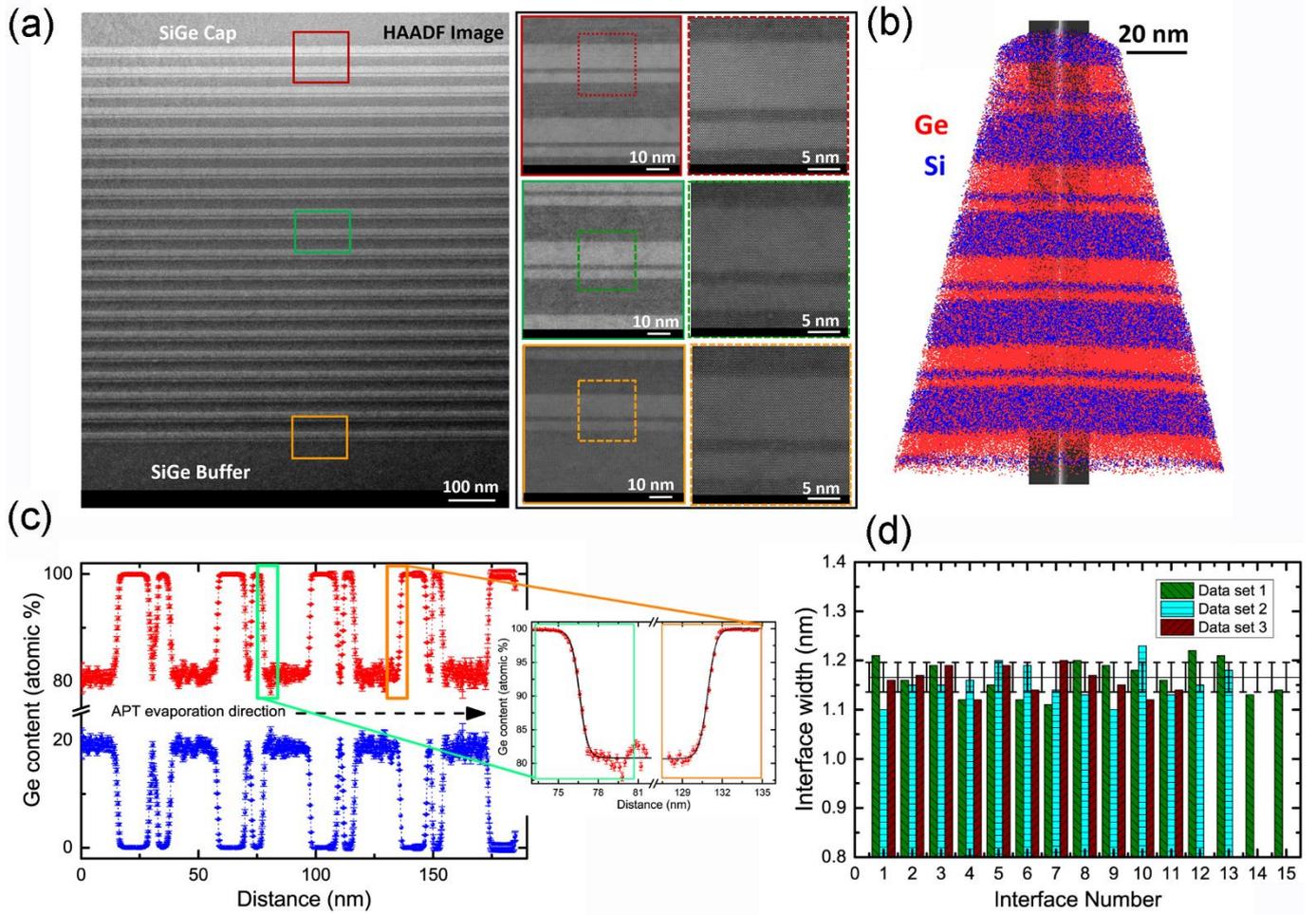

Figure 1: (a) (left) HAADF-STEM image showing all the 20 periods of the Ge/SiGe ACQW, parts of the SiGe capping layer, and the SiGe buffer in the VS. On the right panels, we display images recorded at progressively higher magnification from different sample regions. (b) 3D atom-by-atom reconstruction of the Ge/SiGe ACQW (Data Set 1). (c) 1D concentration profile of Si and Ge (at a fixed bin width of 0.2 nm) along the direction of APT evaporation. Data are extracted from the 30 nm-wide cylinder coaxial to the tip and highlighted in black in panel (b). Inset: Details of a falling (green) and rising (orange) Ge concentration profiles. The black solid lines in the insets represents the fitting function (equation (1), later in the text) used to extract the interface widths. (d) The extracted values of $\mathcal{L}$ for all the interfaces and the three different APT data sets of the sample. The mean value of $\mathcal{L}$ is shown as black horizontal line together with the error bars representing the standard deviation of the measurements on all the interfaces.

Furthermore, the absence of any misfit dislocations and extended defects in the HAADF-STEM images of Fig. 1(a) highlights the good crystalline quality of the sample and the coherent nature of the heterointerfaces.

The atom-by-atom 3D reconstruction of four periods of the sample (Data set-1 of 3) is shown in Fig. 1(b), with the atom evaporation proceeding from the top to the bottom of the tip, *i.e.* in the opposite direction with respect to the growth direction. The second out of the three APT data sets is shown in Fig. S1 of the SM along with a representative mass spectrum. In order to ensure higher accuracy of the atomic scale, only atoms evaporated from a 30 nm wide cylinder coaxial to the tip (see Fig. 1(b)) were considered in the analysis. In fact, the uniformity in the layer-by-layer evaporation is maximum for the atoms located at the center of the hemispherical tip [28].

Figure 1(c) shows the 1D concentration of Si and Ge atoms along the APT evaporation direction. The mean Ge concentration within the SiGe barrier layers was found to be 80.0 % ($\pm 0.6\%$), in perfect agreement with the nominal value. The 1D concentration profile was used to estimate the interface width $\mathcal{L}$. To this aim, in the inset of Fig. 1(c) we show a magnified view of the Ge concentration profile at two Ge→SiGe (green box) and SiGe→Ge (orange box) heterointerfaces, corresponding to the marked green and orange rectangular boxes in the plot of the 1D concentration profile. The raw data was fitted using the error function [11]

$$c(z) = c_0 + d_0 \mathrm{erf}\left[2\sqrt{\ln(2)}\,(z - z_0)\,/\mathcal{L}\right] \qquad (1)$$

where $c_0$ and $d_0$ are an offset and scale parameter, respectively, introduced to achieve the correct concentrations on both side of the heterointerface and $z_0$ defines the position of its center. The interfacial width $\mathcal{L}$ is defined here as the FWHM of the derivative of the fitted concentration profile and corresponds to the interface width over which the concentration changes from 12% to 88% of the plateau value. For the two examples of the rising and falling Ge concentration profiles shown in the inset of Fig. 1(c), the values of $\mathcal{L}$ extracted from the fit were 1.16 and 1.15 nm, respectively.

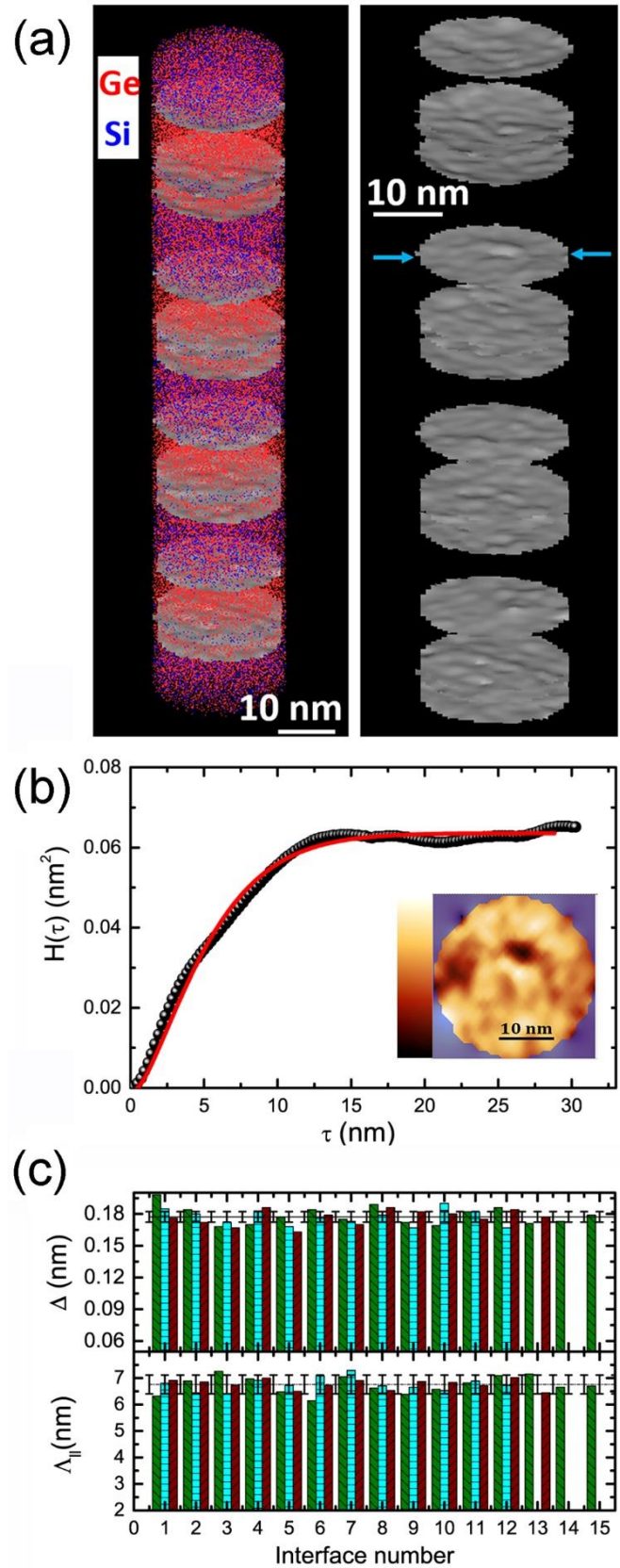

Figure 2: (a) Left: 3D reconstruction of all the atoms located within the black cylinder of diameter 30 nm, shown in Fig. 1(b). Right: Corresponding 90% Ge content iso-compositional surfaces between Ge and SiGe layers. (b) Height-height correlation function $H(\tau)$ as a function of the length of the in-plane vector $\vec{\tau}$ measured on the interface marked by arrow in panel (a). Inset: The color-coded height distribution of the same interface: the color bar range is [-0.38;+0.47] nm. (c) The extracted values of Δ and of $\Lambda_\parallel$ of all the interfaces, for the three different APT data sets. We plotted the mean values of the measurement parameters as black horizontal lines together with the error bars representing the standard deviation of the measurements on all the interfaces.

Figure 1(d) shows the $\mathcal{L}$ values measured from all the interfaces present in the three investigated data sets resulting in an average value of 1.16 nm ($\pm$3.7%). Since often in literature a sigmoid function of the concentration profile is used as alternative fitting profile, we also fitted our data to the function $c_s(z) = c_0 + d_0 / \left[1 + e^{-\frac{(z_0 \pm z)}{L}}\right]$, finding an average value of $L$=0.288 nm. The quantity $4L$ = 1.15 nm corresponds to the length over which the concentration changes from 12% to 88% of the plateau value of the logistic function and can be then compared to the interface width $\mathcal{L}$. Remarkably, the interface width here obtained is only slightly larger than that measured from low Ge content Si/SiGe multi-layers in Ref. [18], using the same technique. The value of $\mathcal{L}$ we report here is also more than a factor ~2.5× smaller compared to those measured in Ge/Si$_{0.2}$Ge$_{0.8}$ multi-quantum well samples grown by reduced pressure CVD and plasma enhanced CVD reactors [29, 30]. This suggests that UHV-CVD might be better suited for realizing sharper interfaces in high-Ge content heterostructures with layer thicknesses of the order of few nanometers. Interestingly, we do not observe any systematic difference between the measured value of $\mathcal{L}$ on the Ge→SiGe and SiGe→Ge interfaces. This observation differs from previous reports on both low- [18, 31] and high-Ge content [30] Si/SiGe heterostructures. In fact, in these early reports larger widths were measured for the interface featuring a Ge content decreasing along the growth direction with respect to the opposite case, as one could expect considering the tendency to surface segregation of Ge atoms during the Si overlayer deposition. We speculate that this discrepancy may be related to specific peculiarities of the UHV-CVD growth process, such the relatively short precursor residential time in the reactor. Finally, as evident from Fig. 1(d), we notice that $\mathcal{L}$ is also independent on both the thickness of the layers and the position within the whole stack. This evidence suggests that the adopted growth temperature of 500°C does not induce relevant intermixing in buried interfaces during the subsequent steps of the growth process [32].

After having characterized the interface width, we now exploit the fully 3D character of our analysis for assessing the roughness Δ and the coherence along both the parallel and axial directions. To this aim, from the 3D atomic distribution shown for instance in the left plot of Fig. 2(a) we preliminary determine the iso-compositional surfaces at Ge concentration of 90%, *i.e.* halfway between the barrier and well compositions (right plot in Fig. 2(a)). Their average positions along the growth direction define a set of $z_0$ values, each of them corresponding to the position of a given heterointerface. This allows the calculation of the height fluctuation $h(\vec{\rho})$ of the 90% iso-compositional surfaces, with respect to their $z_0$, as a function of the in-plane coordinates $\vec{\rho}=(x,y)$. As an example, we report in the inset of Fig. 2(b) the $h(\vec{\rho})$ distribution of one 90% iso-compositional interface (blue arrows in the right plot of Fig. 2(a)). The height-height correlation function $H(\tau)$ is then experimentally determined as the squared difference of $h(\vec{\rho})$ calculated at two points separated by the in-plane vector $\vec{\tau}$ and averaged in the in-plane direction

$$H(\tau) = \langle |h(\vec{\rho}) - h(\vec{\rho}+\vec{\tau})|^2 \rangle_{\vec{\rho}} = 2\langle |h(\vec{\rho})|^2 \rangle_{\vec{\rho}} - 2\langle |h(\vec{\rho})h(\vec{\rho}+\vec{\tau})|^2 \rangle_{\vec{\rho}} \quad (2)$$

In the right-hand side, the first term represents the IFR mean squared roughness $\Delta^2 = \langle |h(\vec{\rho})|^2 \rangle_{\vec{\rho}}$. Under the assumption of in-plane rotation and translation invariance, the second term depends only on the modulus of $\vec{\tau}$, and then can be written as

$$\langle |h(\vec{\rho})h(\vec{\rho}+\vec{\tau})|^2 \rangle_{\vec{\rho}} = \Delta^2 C_\parallel(\tau) \quad (3)$$

where the dimensionless quantity $C_\parallel(\tau)$ is usually referred to as the IFR in-plane correlation function.

In Fig. 2(b) we plot the functional behavior of $H(\tau)$ associated to the exemplificative surface shown in its inset. By assuming a Gaussian decay of the in-plane correlation function[13-17, 19, 33]

$$C_\parallel(\tau) = e^{-(\tau/\Lambda_\parallel)^2} \quad (4),$$

from eq. 4 we can estimate the $\Delta$ and $\Lambda_\parallel$ parameters by fitting the measured $H(\tau)$ to the function $H_{fit}(\tau) = 2\Delta^2 \left[1 - e^{-(\tau/\Lambda_\parallel)^2}\right]$. In this case we obtain $\Delta = 0.18$ nm and $\Lambda_\parallel = 6.98$ nm, respectively.

The values of Δ and $\Lambda_\parallel$ obtained for all the 90% iso-compositional surface across the different APT data sets are highly uniform, as evident in Fig. 2(c). The mean values of Δ and $\Lambda_\parallel$ were found to be 0.18 nm (±5 %) and 6.9 nm (± 4 %), respectively.

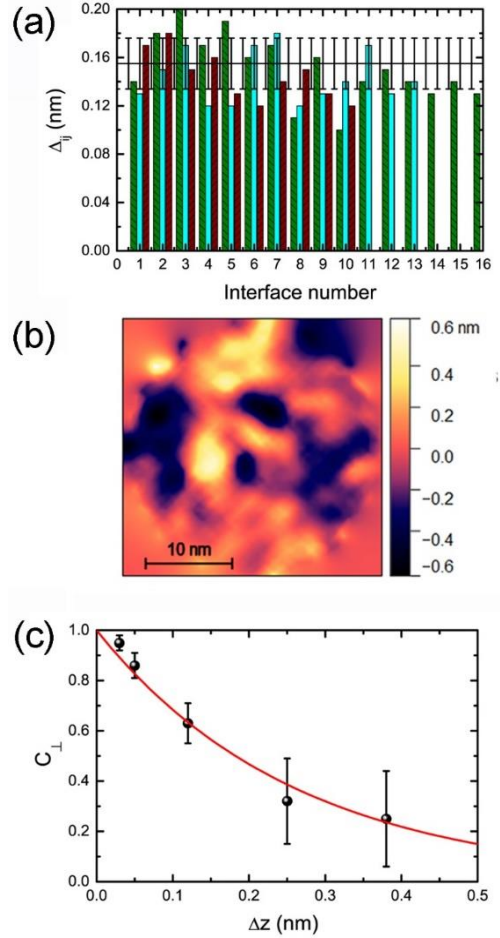

Figure 3: (a) Measured $\Delta_{ij}$ of the $w_{ij}(\vec{\rho})$ differential map with $i = 91\%, j = 89\%$ calculated for all the interfaces, across the different APT data sets. (b): Representative differential map $w_{ij}(\vec{\rho})$ for $i = 91\%$ and $j = 89\%$ iso-compositional surfaces. (c) Data points (spheres) for the axial correlation function fitted to an exponential decay function $C_\perp(\Delta z) = e^{-\Delta z/\Lambda_\perp}$ (line, details in the main text).

On the other end, as already pointed out, for a complete characterization of IFR in diffuse interfaces, the measurement of the roughness correlation length along the growth direction is also required. Therefore, we set up a procedure for the quantification of the axial correlation length $\Lambda_\perp$ from APT data.

To this purpose, we have "sliced" the diffuse interfaces into seven iso-concentration surfaces, defined at different Ge concentrations in the 87 - 93% range (outside this range of concentration values, no meaningful iso-compositional surfaces can be obtained). For each iso-compositional surface we calculated their average positions $z_i = 1 \ldots 7$, obtaining values in perfect agreement with the 1D compositional previously displayed in Fig. 1(c).

Subsequently, the height function $h_i(\vec{\rho})$ associated to each iso-compositional surface was calculated and used to generate the differential maps $w_{ij}(\vec{\rho})$ defined by the equation

$$w_{ij}(\vec{\rho}) = h_i(\vec{\rho}) - h_j(\vec{\rho}) \qquad (5)$$

where the (i,j) indices correspond to iso-compositional surface pairs belonging to the same diffuse interface. Their values were chosen in order to have a data set featuring increasing values of $|\Delta z_{ij}| = |z_i - z_j|$, corresponding to the following couples of iso-compositional surfaces: $(i, j) = (90.5\%, 90\%), (91\%, 90\%), (91\%, 89\%), (92\%, 88\%)$, and $(93\%, 87\%)$. In Fig. 3(a) we show the measured values $\Delta_{ij} = \sqrt{\langle w_{ij}^2(\vec{\rho})\rangle}$ of the rms roughness of the $w_{ij}(\vec{\rho})$ maps for all the interfaces, across the different APT data sets for (i,j) corresponding to (91%,89%). As an example, in Fig. 3(b) we show a representative differential surface maps $w_{ij}(\vec{\rho})$ for this (i,j) pair (the $\Delta_{ij}$ values extracted from all the available $w_{ij}(\vec{\rho})$ maps for other (i,j) values are shown in Fig. S2 in the SM). From these data we can determine the axial correlation function

$$C_\perp(|\Delta z|) = \langle h(\vec{\rho}, z) h(\vec{\rho}, z + \Delta z)\rangle_{\vec{\rho}} / \Delta^2 \qquad (6)$$

To this aim we write the measured $\Delta_{ij}^2$ as

$$\Delta_{ij}^2 = \langle w_{ij}^2(\vec{\rho})\rangle = \langle (h_i(\vec{\rho}) - h_j(\vec{\rho}))^2\rangle = \langle h_i^2(\vec{\rho})\rangle + \langle h_j^2(\vec{\rho})\rangle - 2\langle h_i(\vec{\rho}) h_j(\vec{\rho})\rangle = 2\Delta^2(1 - C_\perp(|z_i - z_j|))$$
(7)

where we assumed that the considered isocompositional surfaces feature the same $\Delta$ value. This relation has been used to plot in Fig. 3(c) the experimental $C_\perp(\Delta z_{ij})$ data points. These values have been fitted to an exponential decay $e^{-\Delta z/\Lambda_\perp}$ function from which we estimated the value $\Lambda_\perp = 0.26$ nm ($\pm 7.5\%$). Remarkably, this quantity is smaller than the mean interface width $\mathcal{L} = 1.16$ nm, meaning that the usually made assumption of fully-correlated roughness within each diffuse interface is not valid [17]. Based on this result, the next section elaborates a general theoretical framework to calculate the impact of a finite value of $\Lambda_\perp$ on the IFR scattering rate.

**B. Theory of IFR scattering rate for diffuse interfaces with finite axial correlation length**

The impact of roughness on electron states in multi-layer structures is usually theoretically addressed in the framework of the perturbation theory, introducing the perturbing potential $\delta V_{IFR}(\vec{\rho}, z)$, which represents the difference between the total 3D potential $V(\vec{\rho}, z)$ and the 1D average potential $\bar{V}(z) = \langle V(\vec{\rho}, z) \rangle_{\vec{\rho}}$, where $\langle \ \rangle_{\vec{\rho}}$ denotes the averaging over the in-plane coordinate $\vec{\rho}$. Hence $\delta V_{IFR}(\vec{\rho}, z)$ represents the fluctuations of the potential that break the in-plane translational invariance [34]. Consequently, $\delta V_{IFR}(\vec{\rho}, z)$ can couple eigenstates of the unperturbed Hamiltonian featuring different in-plane vectors and $z$-dependent envelope functions, i.e. $\varphi_\alpha(z)e^{i\vec{k}\cdot\vec{\rho}}$ and $\varphi_\beta(z)e^{i(\vec{k}+\vec{q})\cdot\vec{\rho}}$.

According to the first-order perturbation theory, the rate of elastic scattering events from the initial state $|\alpha, \vec{k}\rangle$ to any available final state belonging to the $\beta$ subband is given by

$$\hbar \Gamma_{\alpha,\beta,k} = 2\pi \sum_{\vec{q}} |\delta V_{\alpha\beta q}|^2 \delta\left(E_{\alpha\beta} + \frac{\hbar^2 k^2}{2m_\parallel^*} - \frac{\hbar^2 |\vec{k}+\vec{q}|^2}{2m_\parallel^*}\right) \quad (8)$$

where $k$ and $q$ are the modulus of the initial and exchanged momentum, $m_\parallel^*$ is the in-plane effective mass, $E_{\alpha\beta} = E_\alpha - E_\beta$ is the subband energy separation, and $|\delta V_{\alpha\beta q}|^2 = \langle \alpha, \vec{k} | \delta V_{IFR} | \beta, \vec{k}+\vec{q} \rangle$ is the IFR scattering matrix element which is a $k$-independent quantity. Assuming *i)* abrupt interfaces ($\mathcal{L} \to 0$); *ii)* absence of correlation among the different heterointerfaces, *iii)* a Gaussian shape of the two-point correlation function for $h(\vec{\rho})$, this matrix element reads (see Ref. [13, 14])

$$|\delta V_{\alpha\beta q}^{(0)}|^2 = \pi \Delta^2 \Lambda_\parallel^2 e^{-\frac{\Lambda_\parallel^2 q^2}{4}} V_0^2 \sum_\eta |\varphi_\beta(z_0^{(\eta)})|^2 |\varphi_\alpha(z_0^{(\eta)})|^2 \quad (9)$$

In the above equation, the sum extends over the different uncorrelated interfaces located along the growth direction at $z_0^{(\eta)}$ and $V_0$ is the value of the associated band offset which is assumed constant for all the interfaces. The resulting physical picture is that the IFR perturbation can be regarded as a contact potential active at the interface planes and therefore sensible to the local amplitude of the wavefunctions. The square modulus of its matrix element decays exponentially with the square modulus of the exchanged momentum $q$, with a characteristic scale given by $1/\Lambda_\parallel$.

As already mentioned in the introduction, Valavanis *et al.* [17] relaxed the hypothesis of interface abruptness with the scope of investigating SiGe multilayer systems, which typically feature non-negligible interface widths $\mathcal{L}$, in the 0.9-1.5 nm range [18, 31]. Remarkably, the relevance of a proper modeling of diffuse interfaces has recently emerged also in the realm of III-V material systems [11, 35]. In the absence of experimental inputs, Valavanis *et al.* assumed that: *i)* the roughness is fully correlated along growth direction within each diffuse interface, *ii)* the separation between interfaces is much larger than the interdiffusion length, and *iii)* the IFRs of different interfaces are fully uncorrelated (*i.e.* $C_\perp\left(\left|z_0^{(\eta)} - z_0^{(\eta')}\right|\right) = 0$ for $\eta \neq \eta'$). Within these assumptions, the interface broadening was found to have a very limited impact on the IFR scattering rate up to $\mathcal{L} \leq 3$ nm. However, we point out here that not only assumption *i)* may not hold ubiquitously, as directly demonstrated by our experimental findings, but also that the assumptions *ii)* and *iii)* may not be applicable when studying tunneling structures with barrier thickness comparable to the interface width and/or to the axial correlation length. This implies that in many multilayer materials embedded in real-word devices these working hypotheses are not satisfied.

To overcome these limitations and based on the experimental observations outlined in the previous section, we develop here a more general theoretical framework describing the IFR scattering in a generic sequence of diffuse interfaces with a finite correlation length along the axial direction. In presence of diffuses interfaces, the axial correlation can couple the perturbing potential associated to different interfaces and, as a consequence their joint action, cannot be described in terms of a sum of independent scattering sources as in eq. (9). As detailed in the SM, in this general case, the scattering rate depends on the 3D correlation function of the interface roughness. Assuming that this function can be factorized into in an in-plane and an axial part, we obtain

$$|\delta V_{\alpha\beta q}|^2 = \pi \Delta^2 \Lambda_\parallel^2 e^{-\frac{\Lambda_\parallel^2 q^2}{4}} \int dz_1 \int dz_2 \; \varphi_\beta^*(z_1)\varphi_\alpha(z_1)\, \varphi_\alpha^*(z_2)\varphi_\beta(z_2) \frac{\partial \bar{V}}{\partial z}(z_1) \frac{\partial \bar{V}}{\partial z}(z_2) C_\perp(|z_2 - z_1|) \quad (10)$$

where the *z* integrals extend over the whole heterostructure stack. As such, we notice that this equation, being formulated in terms of a generic potential profile $\bar{V}(z)$ defined throughout the sample, can be applied without artificially describing $\bar{V}(z)$ as a set of different contributions coming from partially overlapping "*blurred*" interfaces [36].

To gain insight on the impact of the correlation effect due to a finite value of $C_\perp$, we calculate the scattering rate $\Gamma_{\alpha\beta k}$ from an initial state $|\alpha, k\rangle$ to a different subband, obtained summing over all available wavevectors $\vec{k_f} = \vec{k} + \vec{q}$ in the final subband:

$$\Gamma_{\alpha\beta k} = \frac{m_\parallel^*}{\hbar^2} \int_0^\pi d\theta\, e^{-\frac{\Lambda_\parallel^2 q^2(\theta)}{4}} |\delta V_{\alpha\beta q}|^2 \delta\left(E_{\alpha\beta} + \frac{\hbar^2 k^2}{2m_\parallel^*} - \frac{\hbar^2 |\vec{k}+\vec{q}|^2}{2m_\parallel^*}\right) \quad (11)$$

where we have taken into account the energy conservation to express the modulus of the exchanged wavevector as a function of the scattering angle $\theta$ only.

We now shall disregard the correlation possibly existing between different interfaces, considering one heterointerface only or, alternatively, we can suppose that the axial correlation function $C_\perp$ is negligible when calculated at the minimum heterointerface distance featured by the multilayer stack. Under the additional hypothesis that the variations of the wavefunctions at the interface width scale are small, it can be easily proven (see SM) that $\Gamma_{\alpha\beta k} = F \Gamma_{\alpha\beta k}^{(0)}$ where

$$F = \frac{1}{V_0^2} \int dz_1 \int dz_2\, \frac{\partial \bar{V}}{\partial z}(z_1) \frac{\partial \bar{V}}{\partial z}(z_2) C_\perp(|z_2 - z_1|) \quad (12)$$

and $\Gamma_{\alpha\beta k}^{(0)}$ is the scattering rate calculated according to eq. (9). Aiming at a quantitative estimation of the $F$ value, we calculate the right-end side of the above equation in the case of an exponentially decaying axial correlation function and assuming an error-function profile for the diffuse interface, as suggested by the fit of our APT experimental data. In these conditions, we obtain that $F$ is controlled by the dimensionless parameter $\frac{\mathcal{L}}{\Lambda_\perp}$ according to

$$F = \exp\left(\frac{\mathcal{L}^2}{16\ln(2)\Lambda_\perp^2}\right)\left[\text{erf}\left(-\frac{\mathcal{L}}{4\ln(2)\Lambda_\perp}\right) + 1\right] \quad (13)$$

Interestingly, for our measured values of $\mathcal{L} = 1.16$ nm and $\Lambda_\perp=0.26$ nm we find F=0.35, meaning that IFR scattering rate is *reduced* by a factor ~3× with respect to the value predicted for an abrupt interface with the same rms roughness.

We can also define an equivalent rms roughness for abrupt interfaces a $\Delta_{eq}^2 = \Delta^2 F$. The value F=0.35 calculated in this work gives an equivalent rms roughness $\Delta_{eq} = 0.106$ nm instead of $\Delta = 0.18$ nm.

This reduced effectiveness of the IFR potential can be explained considering that the interface does not behave as a single scattering center but, instead, acts as a collection of different, only partially correlated, scattering centers. We notice that, in line with Eq. (12), $F\sim 0$ when $\Lambda_\perp \ll \mathcal{L}$. On the contrary, for $\Lambda_\perp \gg \mathcal{L}$, F ~ 1 and, consequently, the abrupt interface limit is recovered.

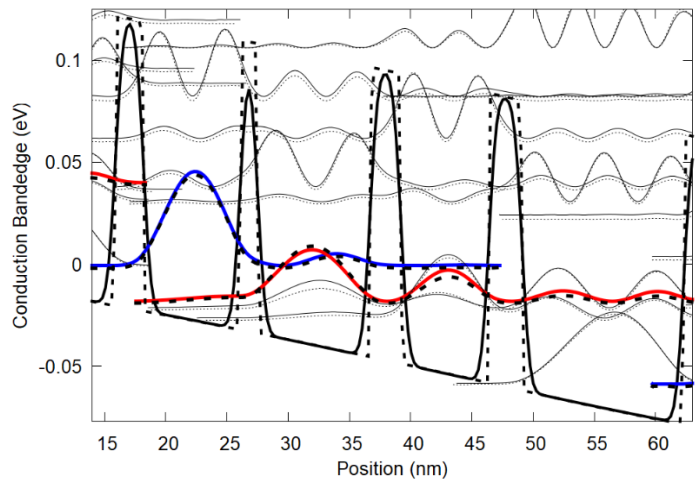

Figure 4: L-valley band-edge and electronic states of a THz n-type Ge/Ge$_{0.75}$Si$_{0.25}$ QCL with abrupt interfaces (dashed lines) and with interdiffusion (solid lines). The squared wavefunctions associated to the upper (ULL, blue) and lower (LLL, red) laser levels are highlighted.

**Impact of IFR scattering rate on the performances of a QCL device**

As previously discussed in the introduction, a thorough theoretical and experimental understanding of heterointerfaces can greatly help in designing and modeling innovative devices. As a relevant example, we discuss here the impact of the IFR scattering on the performances of a n-type Ge/SiGe quantum cascade laser grown on top of a Si(001) substrate, featuring a composition profile quite similar to the one of the ACQW samples here investigated. This choice is motivated by the fact that QCL systems are a prototypical case-study to assess the role of IFR scattering [6, 37, 38], owing to the inherently large number of heterointerfaces featured by this class of devices, and by the technological relevance of a Si-based THz source, suitable for room temperature operation [20, 21].

In Fig. 4, we show the L-point band edge energy profile together with the relevant states, calculated in the envelope function approximation, considering both abrupt (dashed lines) and diffuse (solid lines) interfaces for which, guided by our experimental results, we assumed $\mathcal{L} = 1.16$ nm. From Fig. 4, we predict that this moderate degree of interdiffusion has a very limited impact on the wave functions of the low-energy states controlling the QCL carrier dynamics. Actually, in order to recover the same optical properties predicted for the abrupt heterointerface condition, the effect of interface broadening on the gain can be compensated by an appropriate fine-tuning of the layers thickness and composition (see Fig. S3 in the SM).

In Fig. 5 (a), we report the IFR scattering rate from an initial state in the upper laser level (ULL) subband with zero in-plane momentum to the lower laser level (LLL) subband calculated as a function of $\Lambda_\parallel$, setting the values $\mathcal{L} = 1.16$ nm, $\Lambda_\perp = 0.26$ nm and $\Delta = 0.18$ nm obtained by APT data. The scattering rate tends to zero both when $\Lambda_\parallel \to 0$ and $\Lambda_\parallel \to \infty$ with a maximum for $\Lambda_\parallel \sim 5.6$ nm, unfortunately quite close to the value of 6.9 nm measured in our ACQWs samples (vertical dashed line). To understand the reason for the occurrence of a peak in Fig.5(a) at ~5.6 nm, we remind that for an in-plane mass of 0.32 $m_0$ (associated to the effective mass tensor of Ge(001) at the L point) and an energy separation between the two laser levels of 16 meV, the modulus of the wave vector exchanged in the IFR elastic scattering process is $q \sim 0.35$ nm$^{-1}$. Considering now the functional dependence on $\Lambda_\parallel$ in Eq.(10), the scattering rate is expected to have a maximum when $2/\Lambda_\parallel = q$, corresponding to $\Lambda_\parallel = 5.7$ nm when the exchanged momentum $q$ is 0.35 nm$^{-1}$, in great agreement with the observed peak value in Fig. 5(a). The results of Fig. 5(a) suggest that a reduction of the IFR scattering rate can be achieved by mismatching the $q$ and $2/\Lambda_\parallel$ scales: to this aim one either can act on the laser level separation, or should modify the in-plane correlation length at the growth stage, which is clearly a more challenging task.

The dependence of the IFR scattering rate between the two laser levels on the axial correlation length obtained from our model, setting $\mathcal{L} = 1.16$ nm, $\Lambda_\parallel = 6.9$ nm and $\Delta = 0.18$ nm as experimentally determined, is displayed by the blue curve in Fig. 5(b); the dashed vertical line has been drawn at $\Lambda_\perp = 0.26$ nm as in our sample. For comparison, the horizontal lines represent the IFR scattering rate obtained using the abrupt interface model of equation (9) (red line) and the perfectly correlated diffuse interface model developed by Valavanis *et al*. in Ref. [17] (green line), both neglecting possible coupling among different interfaces. Notice that the deviation of the IFR scattering rate obtained from the Valavanis model with respect to that predicted by the abrupt interface model is due to the merging of the rising and falling concentration profiles in the diffuse interfaces defining the tunnel barrier located between the ULL and the LLL, which lowers the potential barrier height and hence slightly suppresses the scattering rate with respect to the atomically sharp case (see dashed and continuum band edge profiles in Fig. 4). For the QCL structure considered here, this deviation emerges despite $\mathcal{L} < 3$ nm, thus in a range where according to Ref [17] the impact of interface broadening on the IFR scattering rate should be negligible. This apparent contradiction is due to the fact that our QCL structure features thin tunneling barriers with a thickness in the order of $\mathcal{L}$, and thus does not satisfy the assumption of interface separations larger than $\mathcal{L}$ considered in Ref [17].

Our model (blue curve) indicates that the IFR scattering rate, in the range explored in Fig. 5(b), has a non-monotonic behavior, being an increasing (decreasing) function for small (large) values of $\Lambda_\perp$, with a maximum value achieved at $\Lambda_\perp \sim 1.2$ nm. To understand this functional behavior, we first notice that a multilayer system is characterized by two length scales: *i)* the interface width $\mathcal{L}$ and *ii)* the minimum distance ($d_{min}$) separating two distinct interface planes featuring non negligible amplitudes of both the interacting wave functions. In our QCL design and for the considered states, $d_{min}$ corresponds to the thickness of the tunneling barrier separating the two wells which host the laser levels ($d_{min} = 1.4$ nm). This value is approximately equal to $\mathcal{L} = 1.16$ nm (see Fig. 4). Consequently, the two length scales in our case are practically coincident. To understand the rising part of the blue curve, we notice that, for $\Lambda_\perp \ll d_{min}$, the coupling among different interfaces is negligible and, therefore, the double spatial integral of Eq. (10) can be split in a discrete sum of terms, each representing the contribution from an isolated interface, analogously to the case of equation (9). In other words, in this regime $C_\perp(|z_2 - z_1|)$ is not negligible only when $z_1$ and $z_2$ range in a neighbor of the same interface. Consequently, the spatial derivative $\frac{\partial V}{\partial z}(z_1)\frac{\partial V}{\partial z}(z_2)$ and the two wavefunctions $\varphi_\alpha(z_1)\varphi_\alpha^*(z_2)$ and $\varphi_\beta(z_1)\varphi_\beta^*(z_2)$ products entering the integral in eq. (10) have positive values.

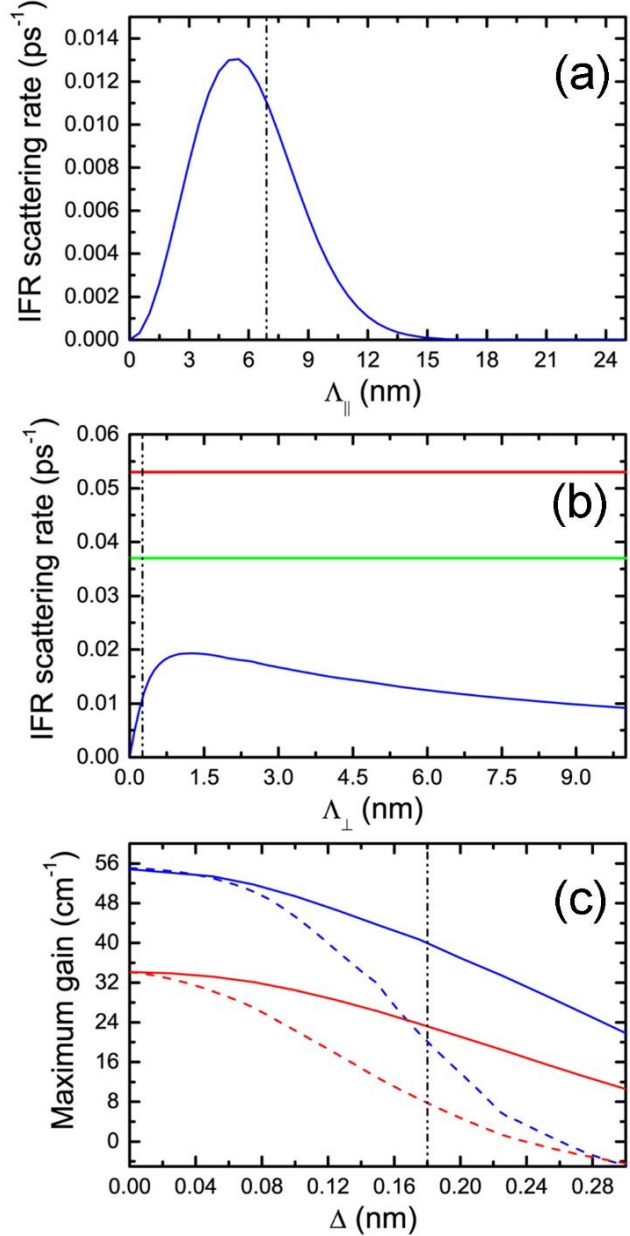

Figure 5: (a, b) IFR scattering rate (blue curve) between the QCL laser levels calculated with our model (a) as a function of in-plane correlation length $\Lambda_\parallel$ setting $\mathcal{L} = 1.16$ nm, $\Lambda_\perp=0.26$ nm and $\Delta=0.18$ nm; (b) as a function of axial correlation length $\Lambda_\perp$ setting $\mathcal{L} = 1.16$ nm, $\Lambda_\parallel =6.9$ nm and $\Delta=0.18$ nm. In panel (b), reference values corresponding to the IFR scattering rate calculated with the same values of $\Lambda_\parallel$ and $\Delta$ using the abrupt interface model ($\mathcal{L} = 0$) and the Valavanis model for diffuse interfaces [17] ($\mathcal{L} = 1.16$ nm) are displayed in red and green, respectively. (c) Maximum QCL gain as a function of the rms interface roughness $\Delta$ for lattice temperatures of 10 K (blue) and 300 K (red), by taking into account interdiffusion (solid lines) or neglecting it (dashed). The calculations were performed with $\mathcal{L} = 1.16$ nm, $\Lambda_\parallel=6.9$ nm, and $\Lambda_\perp=0.26$ nm. In the three panels, vertical lines mark the values measured by APT on our samples ($\Lambda_\parallel =6.9$ nm, $\Lambda_\perp=0.26$ nm, $\Delta=0.18$ nm).

This condition, together with the monotonic increase of $C_\perp(|z_2-z_1|)$ with $\Lambda_\perp$, produces the rise of the IFR scattering rate for $\Lambda_\perp \ll d_{min}$. When $\Lambda_\perp$ approaches $d_{min}$, the correlation between interfaces starts to become relevant, since mixed terms in the double spatial integral, associated to the coupling of different interfaces, are no more suppressed by the $C_\perp(|z_2-z_1|)$ factor. Interference effects between the different interfaces then come into play, with a positive or negative contribution that in general depends on the particular potential shape and on the considered wavefunctions. In the case of the QCL of Fig. 4, considering the IFR scattering between the ULL and the LLL subbands, the interference effects are mainly due to the interaction between the two interfaces that define the tunneling barrier at a position of ~27 nm. In such case, when $z_1$ and $z_2$ belong to the two consecutive interfaces, the product $\frac{\partial V}{\partial z}(z_1)\frac{\partial V}{\partial z}(z_2)$ is a negative number, while the $\varphi_\alpha(z_1)\varphi_\alpha^*(z_2)\varphi_\beta(z_1)\varphi_\beta^*(z_2)$ product has a positive value. This explains why an increase of the axial correlation $\Lambda_\perp$ beyond 1.2 nm triggers a decrease of the IFR rate. Note that the IFR scattering rate predicted by our model never reaches the limiting value of the Valavanis model (green curve) since the two assumption of Valavanis model *i.e. i)* $\Lambda_\perp \ll d_{min}$ (uncoupled interfaces) and *ii)* $\Lambda_\perp \gg \mathcal{L}$ (perfect correlation within each interface) cannot be simultaneously fulfilled in our QCL design for which $\mathcal{L} \approx d_{min}$. Remarkably, at the value $\Lambda_\perp = 0.26$ nm experimentally evaluated on the grown structure, we obtain a IFR scattering rate lower by a factor ~3× with respect to the value predicted by the Valavanis model. It is interesting to note that this suppression factor matches the analytical estimation of the F value in eqs. (12-13) which resulted in a value of 0.35 for the IFR parameters measured on our sample.

The main message of Fig. 5(b) is thus that the parameter $\Lambda_\perp$ has a crucial effect on the IFR rate and should be properly considered when designing tunneling heterostructure devices. In fact, two effects are simultaneously affecting the scattering in tunneling structures: the non-perfectly correlated roughness within the single interface and the coupling of the IFR potential associated to different interfaces. Therefore, for fixed $\Lambda_\perp$, changes in the multilayer thicknesses and/or in the wavefunction profile can result in relevant variation of the IFR scattering rate.

Finally, for the QCL structure of Fig. 4, we quantitatively compare the maximum gain as a function of $\Delta$ calculated using the generalized model of IFR scattering described in this paper (solid lines) with that predicted by the abrupt uncorrelated interface approach (dashed lines), as done in Refs. [21] and [39] [Fig. 5(c)]. For each model, the gain is estimated for lattice temperatures of 10 K (blue) and 300 K (red). Again, the calculations were performed setting the experimentally determined interface parameters for $\mathcal{L}$, $\Lambda_{\parallel}$ and $\Lambda_{\perp}$. We observe that, for the case of our model with diffuse interfaces, the gain is much more robust against an increase of $\Delta$, due to the diminished detrimental impact of the IFR scattering rate, as pointed out when discussing Fig. 5(b). Interestingly, for the APT measured value $\Delta = 0.18$ nm (vertical dashed line), our model predicts gain values more than doubled with respect to the abrupt case. To appreciate the relevance of this result we observe that for a realistic value of optical losses (*i.e.* waveguide plus facet losses) of ≈20 cm$^{-1}$ [21] laser action at room temperature can be expected.

## IV. CONCLUSIONS

By exploiting the elemental-resolved resolution of APT at the atomic scale, we performed a complete characterization of the interface properties in UHV-CVD grown Ge/SiGe quantum heterostructures, obtaining the full set of interface parameters, including width, rms roughness, in-plane and axial correlation of the heterointerfaces. Our results demonstrate the excellent interface quality in terms of small interface width ($\mathcal{L} = 1.16$ nm) and interface rms roughness ($\Delta = 0.18$ nm). We also found that the interface roughness is correlated along the growth axis on a length ($\Lambda_{\perp} = 0.26$ nm) smaller than the interfacial width $\mathcal{L}$. Such a partial coherence of roughness along the growth direction was neglected by available models of the IFR scattering. To obtain a realistic picture consistent with the measured structural data, we therefore developed a generalized theory for IFR scattering of carriers in semiconductor heterostructures featuring a non-negligible interface width. With our measured value of the axial correlation length, this model predicts a reduction of IFR scattering by a factor ~3×, as compared to that corresponding to an abrupt interface featuring the same rms IFR value and a perfect axial correlation within the interface width.

Finally, by using NEGF simulations, we predicted the impact of such reduced IFR scattering rate on the maximum gain of a Ge/SiGe-based quantum cascade laser structure, featuring the same material parameters measured in our ACQW structures. The estimated maximum gain was found to be double of that obtained using oversimplified models. We underline that the technique of extracting the interfacial parameters demonstrated in this work in the case of the GeSi heterostructures can be extended to a variety of other material systems like the III/V multi-quantum wells, group IV-dielectric interfaces, III/V-dielectric interfaces, provided clean and high quality APT maps of these interfaces are achieved. Furthermore, our modeling of scattering by non-abrupt interfaces can be extended to the calculation of roughness-limited mobilities in two-dimensional electronic systems.

**AKNOWLEDGMENTS**

This work is supported by the European Union Horizon 2020 program under grant No.766719 -FLASH Project. The work carried out in Montreal was supported by Canada Research Chairs, NSERC (CRD, SPG, and Discovery Grants), Defence Canada (Innovation for Defence Excellence and Security, IDEaS) and PRIMA Québec.

Fruitful discussions with Giacomo Scalari, Jerome Faist, and Douglas J. Paul are gratefully acknowledged.